# An Analysis of Centrality Measures for Complex and Social Networks

Felipe Grando
Institute of Informatics
UFRGS
Porto Alegre – RS, Brazil
fgrando@inf.ufrgs.br

Diego Noble
Institute of Informatics
UFRGS
Porto Alegre – RS, Brazil
dvnoble@inf.ufrgs.br

Luis C. Lamb
Institute of Informatics
UFRGS
Porto Alegre – RS, Brazil
lamb@inf.ufrgs.br

*Abstract*—Measures of complex network analysis, such as vertex centrality, have the potential to unveil existing network patterns and behaviors. They contribute to the understanding of networks and their components by analyzing their structural properties, which makes them useful in several computer science domains and applications. Unfortunately, there is a large number of distinct centrality measures and little is known about their common characteristics in practice. By means of an empirical analysis, we aim at a clear understanding of the main centrality measures available, unveiling their similarities and differences in a large number of distinct social networks. Our experiments show that the vertex centrality measures known as *information*, *eigenvector*, *subgraph*, *walk betweenness* and *betweenness* can distinguish vertices in all kinds of networks with a granularity performance at 95%, while other metrics achieved a considerably lower result. In addition, we demonstrate that several pairs of metrics evaluate the vertices in a very similar way, i.e. their correlation coefficient values are above 0.7. This was unexpected, considering that each metric presents a quite distinct theoretical and algorithmic foundation. Our work thus contributes towards the development of a methodology for principled network analysis and evaluation.

## I. INTRODUCTION[*]

Complex networks are ubiquitous in various technological, social and biological domains. Several computer science domains make use of complex and social networks, which can be conceptualized as lying at the intersection between artificial intelligence (AI), graph theory and statistical mechanics, displaying a truly multidisciplinary nature [5][14]. Complex network analysis is fundamental to the understanding and modeling of human, social and economic relationships [14]. Both the description and categorization of natural and human-made structures using complex networks lead to the important question of how to choose the most appropriate metrics and evaluations of structural properties.

While such a choice should reflect specific interests and applications, unfortunately there is no general model, formal procedure or methodology for identifying the best measurements for a given network. In addition, the large number of metrics and their respective variations are often related despite the fact that each one of them consider distinct ideas for measuring graph properties. Ultimately, one has to rely on unwarranted intuition or limited knowledge about the problem to decide which metric is the most suitable for an application and to interpret it properly.

Centrality measures can be viewed as a mathematical heuristic applied to network analysis to identify important elements of the networks through their structural properties (i.e. topology features). Nonetheless, there is no widely accepted formal definition of centrality in the context of social complex networks. This fact contributed to the proposal of many divergent centrality measures with respect to different network concepts, such as importance, power, authority, control, independency and influence. Moreover, each centrality measure is capable of evaluating different aspects of a given network depending on the application context and analysis approach.

Furthermore, there is no formal procedure to guide the choice of centrality measures for a given application. However, some works have evaluated important characteristics with respect to the use of existing metrics (see e.g. [8][9][18][35]). Unfortunately, the relevance of their work in real-world applications is restricted to specific domains as they use a small number of experimental samples and centrality measures or restrict their analysis to specific kinds of applications.

Considering the limitations presented above and aiming at solving them, we shall analyze and study the relationship among metrics and detail their characteristics, relating them with networks' structural properties. Our results suggests that the application of centrality measures simultaneously can lead to limited or similar results due their high correlation and redundancy. Furthermore, we showed that the structural properties of the networks do not significantly affect the centralities granularity while they do have a considerable impact on their correlation.

The remainder of the paper is organized as follows. In Section II, we introduce the centrality measures and the complex network models while also explain briefly the eight metrics and six models used in our experiments and analysis. In Section III, we describe our experimental methodology while in Section IV we present our main results. Section V summarizes our conclusion, points out our most important contributions and presents guidelines for future research.

---

## II. ON NETWORKS AND MEASURES

Centrality measures aim at quantifying how important an element of the network is by relying only on the structural pattern of the network. The centrality values are also used frequently as a ranking or identification system [1] and can be effectively used for finding and evaluating subgoals in multi-agent systems [32].

The vertex centrality measures have been used by many works in different areas, including: strategic network formation [6], game theory [23], social behavior [13], transportation [20], influence and marketing [22], communication [37], scientific citation and collaboration [27], communities [29], group problem-solving [31].

Each centrality measure capture a distinct idea of vertex local/global importance in a given network or graph. *Betweenness* and *walk betweenness* refer to an idea of control (important vertices are the ones across many paths and have a potential control over these paths). *Closeness*, *eccentricity* and *information* capture the idea of independency (important vertices are the ones closer to all others, therefore most independent from the others). While, *degree*, *eigenvector* and *subgraph* quantify visibility (important vertices are the ones most noticeable and/or directly involved in the network's substructures).

Table I presents the main characteristics of the eight selected metrics' and their references for detailed information about their underlying ideas and definitions. Notice that the metrics chosen are applicable only in undirected and unweighted graphs. However, there are variations that extend their use to more general kinds of networks that we do not discuss in this work. It is also important to highlight that the time and space complexity vary over each metric and can be a relevant issue when working with massive networks/graphs. Thus, there are important algorithms to reduce the computational costs that are not presented in this paper.

We recall that social networks are usually modeled by graphs where each vertex represents a node of the network and each edge may represent any kind of relationship between such nodes. The overall structure of a network has consequences not only over individual members, but also over the entire group. Furthermore, structural properties of a network may extend well beyond individual behaviors and social roles. Assessing the quality of relations between entities and understanding connection patterns has generated much interest and research in various disciplines [11].

Studies of real complex networks have shown that they have several relevant properties such as "small-world" effect (low diameter), scale-free effect (heavy-tailed degree distribution) and community organization (high global clustering coefficients), among others. Such studies proposed distinct models capable of generating synthetic random networks mapping each one of these characteristics. Their objective is to understand how real networks are organized and to provide tools for network generation and investigation, valuable for many researches [10].

The six complex network models used in our experiments are briefly summarized in Table II.

TABLE I. CENTRALITY MEASURES SUMMARY

| Centrality | Formulae | | Underlying Idea |
|---|---|---|---|
| Betweenness ($C_b$) [17] | $C_b(p_k) = \sum_{i=1}^{n} \sum_{j=i+1}^{n} \frac{g_{ij}(p_k)}{g_{ij}}$ | $g_{ij}(p_k) = $ number of geodesics between $p_i$ and $p_j$ that contains $p_k$ | Communication control and frequency of arrival |
| Closeness ($C_c$) [17] | $C_c(p_k) = \frac{1}{\sum_{i=1}^{n} d(p_i, p_k)}$ | $d(p_i, p_k) = $ geodesic distance from $p_i$ to $p_k$ | Independency, efficiency and time-until-arrival |
| Degree ($C_d$) [17] | $C_d(p_k) = \sum_{i=1}^{n} a(p_i, p_k)$ | $a(p_i, p_k) = \begin{cases} 1 \text{ if } p_i \text{ and } p_k \text{ are adjacent} \\ 0 \text{ otherwise} \end{cases}$ | Visibility and communication activity |
| Eccentricity ($C_x$) [19] | $C_x(p_k) = \frac{1}{\max_{i \in V} d(p_i, p_k)}$ | $d(p_i, p_k) = $ geodesic distance from $p_i$ to $p_k$ | Related to *closeness* but considers only the largest geodesic path |
| Eigenvector ($C_e$) [7] | $C_e(p_k) = E_{[k]}^{+\infty}$ where $E^{+\infty} = \sum_{it=1}^{+\infty} \frac{E^{it}A}{\sum_{i=1}^{n} E_{[i]}^{it-1}}$ | $A = $ adjacency matrix with unitary values on the main diagonal $E^0 = $ vector complete with unitary values | The power of your friends is your power of influence |
| Information ($C_i$) [34] | $C_i(p_k) = \frac{1}{b_{kk} + \frac{T-2R}{n}}$ $T = \sum_{j=1}^{n} b_{jj}$ and $R = \sum_{j=1}^{n} b_{ij}$ for any fixed $i$ | $B = (D - A + U)^{-1}$ $D = $ diagonal matrix with degree values $A = $ adjacency matrix $U = $ matrix having all unitary elements | Related to *closeness* but considers all paths |
| Subgraph ($C_s$) [16] | $C_s(p_k) = \lim_{e \to +\infty} \frac{\left(\frac{(A^e)_{kk}}{e!}\right)}{\sum_{i=1}^{n} A_{ii}}$ | $A = $ adjacency matrix | Involvement and contribution |
| Walk Betweenness ($C_w$) [26] | $C_w(p_k) = \sum_{i=1}^{n} \sum_{j=i+1}^{n} I_{ij}(p_k)$ $I_{kj}(p_k) = 1$ and $I_{ik}(p_k) = 1$ $I_{ij}(p_k) = \frac{1}{2}\sum_{t=1}^{n} A_{kt}\|T_{ki} - T_{kj} - T_{ti} + T_{tj}\|$ for $k \neq i, j$ | $T = \begin{pmatrix} (D-A), \\ \text{remove a single row and} \\ \text{its corresponding column} \end{pmatrix}^{-1}$, add back the removed row and column with values zero $D = $ diagonal matrix with degree values $A = $ adjacency matrix | Related to *betweenness* but considers all paths |

TABLE II. COMPLEX NETWORK MODELS SUMMARY

| Network Model | Parameters in Experiments |
|---|---|
| Networks with Community Structure ($M_{cs}$) [30] | $p_c$ = probability of each vertex to belongs to each community {0.1}. <br> $p$ = edge probability between two vertices belonging to a common community {0.5, 0.7}. <br> $c$ = number of communities {$n/10, n/20, n/50$}. <br> $n$ = number of vertices {100, 500}. |
| Simple Random Graphs ($M_{er}$) [15] | $p$ = probability of connecting each pair of vertices {0.1, 0.3, 0.5}. <br> $n$ = number of vertices {100, 500}. |
| Geographical Models ($M_{gr}$) [12] | $p_{ij}$ = probability that vertices $i$ and $j$ are connected {$k^{-s_{ij}}$}. <br> $s_{ij}$ = distance between vertices $i$ and $j$ {$\left\lvert \left\lfloor \frac{i}{\sqrt{n}} \right\rfloor - \left\lfloor \frac{j}{\sqrt{n}} \right\rfloor \right\rvert + \left\lvert (i \bmod \sqrt{n}) - (j \bmod \sqrt{n}) \right\rvert$}. <br> $k$ = variable used in the $p_{ij}$ equation {1.2, 1.5, 2}. <br> $n$ = number of vertices organized in a determined space {100, 500}. |
| Scale-Free Networks ($M_{sf}$) [5] | $k$ = initial number of fully connected vertices and the number of edges added with each new vertex {2, 3, 5}. <br> $n$ = final number of vertices {100, 500}. |
| Small-World Model ($M_{sw}$) [36] | $p$ = probability of changing each connection (relinking) {0.1, 0.3, 0.5}. <br> $k$ = initial number of nearest neighbors which a vertex is connected {4, 8, 16}. <br> $n$ = number of vertices connected in a ring structure {100, 500}. |
| Kronecker Graphs ($M_{kg}$) [24] | $P$ = square matrix of order two with parameters estimated for different kinds of networks: social {Email-Inside, Epinions}, information/citation {Blog-Nat06All}, Web {Web-Notredame}, Internet {As-Newman, As-RouteViews} and biological {Bio-Proteins}. <br> $n$ = number of vertices {$2^7$=128, $2^9$=512}. |

## III. EXPERIMENTAL METHODOLOGY

We have chosen six complex network models with a hundred samples for each possible combination of the selected parameters (presented in Table II). The parameters were chosen following a preliminary experimental setup that verified the main properties of the networks generated by each model. These models provide a method capable of generating networks that are accurate approximations of real networks properties and behavior [28].

The network models were used to provide several samples to achieve a good generalization of our hypotheses and results so that they were statistically relevant.

A sample network generated by each complex model is presented in Figure 1. Every network shown in Figure 1 has 100 vertices, except for the $M_{kg}$ sample, which has 128 vertices. The algorithm Force Atlas 2 [21] from the Gephi tool (*http://gephi.github.io/*) was used to organize the visualization of the networks, and the vertices' sizes are proportional to their degree.

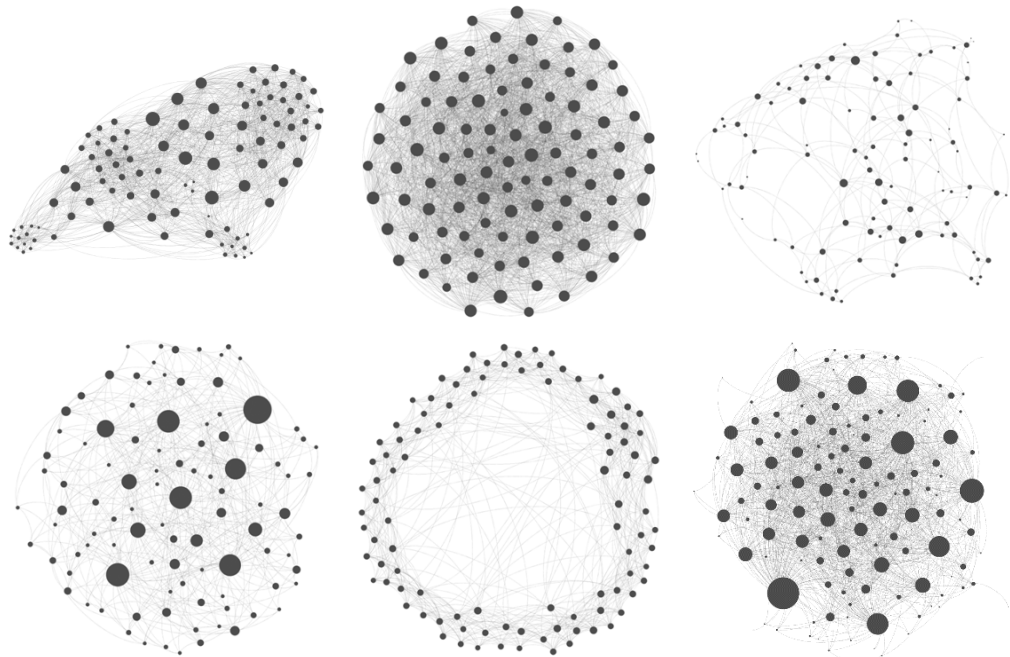

Fig. 1. Sample Networks ($M_{cs}$, $M_{er}$, $M_{gr}$, $M_{sf}$, $M_{sw}$, $M_{kg}$ respectively from top-left to bottom-right)

We have not used larger networks due to the time complexity of some centrality measure algorithms (e.g., *walk betweenness* is cubic in the number of vertices rendering its computation unfeasible). Even simpler metrics, such as *closeness*, are quite demanding for big networks. It can take several weeks to compute just one simple metric in a single network with millions of vertices with a high performance computer (as seen in [11]). Besides, our results and methods could be easily extended to arbitrary size networks considering that the set of network's structural properties remains nearly the same in larger networks' sizes.

We have also used non-isomorphic connected graphs ($N_{ni}$). Two graphs are non-isomorphic if there is no possible edge permutation that transform a given graph into the other. We selected all $N_{ni}$ of six (112 graphs) and seven vertices (853 graphs) retrieved from a public dataset available at *http://cs.anu.edu.au/~bdm/~data/graphs.html*. These graphs are useful to illustrate possible extreme configurations for centrality values. Some of these are very unlikely in random generative models. The sizes (six and seven vertices) were chosen because they allow variability and they can be subject to experimental analyses (e.g., there is a total of 11,716,571 non-isomorphic connected graphs of ten vertices). Our final experimental setup contained 7,165 synthetic networks.

## IV. EXPERIMENTAL ANALYSIS

The analysis and discussion of our experimental results are divided into two groups. The first presents the correlation of centrality measures and the second describes the metrics' granularity (percentage of distinct values).

### A. Centrality Measure Similarity

The analysis of our experiments starts with the centrality measures' correlation to check their similarity. The comparison between the measured centrality values via different centrality measures can unveil relevant information about a network structure. We can assess a distinct role for a given vertex in a network considering that each metric gives a different meaning of vertex importance. Therefore, the correlation among metrics gives an idea of the relation between such distinct roles.

We used the *Kendall Tau-b* rank correlation coefficient between every combination of metrics to estimate the metrics' relationship. This coefficient evaluates the degree of similarity between two sets of ranks given to the same set of objects. Kendall's correlation is especially useful for centrality measures because varying normalization and distribution do not affect it. In addition, centrality measure values are used frequently as ranking factors where the absolute value itself is irrelevant. The same analytical approach was used by other works in similar areas [4][33].

The mean Kendall rank correlation coefficients between each pair of metrics are summarized in Table III. All values presented in the table match the real expected value with 0.01 confidence interval (above and below) with 99% chance. The correlation values above or equal 0.8 are highlighted in bold.

We see that just a few pairs of centralities have low correlation (excluding those with *eccentricity*), which means that despite all centrality differences, they all have a high common agreement when ranking vertices of a network by their centrality in all kinds of networks tested. This is a strong indicative that the simultaneous use of some pairs of metrics is quite redundant and will provide almost certainly fruitless or nearly equivalent results. Therefore, the simpler metrics are suitable for most applications, being easier to interpret and having lower computational complexity.

*Eccentricity* presented the lowest correlation values by far, mainly because it evaluates many vertices as being equally central. The Kendall *tau-b* rank correlation accounts positively for tied values only if both sequences present the same elements with equivalent ranks, penalizing them otherwise.

*Eigenvector* and *subgraph* centralities acquired a perfect correlation coefficient (1.00) in most networks but, at the same time, they presented low correlation in some rare cases (below 0.5). This emphasizes that both of them are so similar that choosing one metric over the other does not have any impact in the results. It also means that the idea of centrality behind each metric (Table I) are equivalent in a practical sense.

We also notice that the correlation values among metrics vary according to the generative model. In the networks generated by $M_{sw}$ and by $M_{sf}$, the overall correlation values for all metrics where much lower (23% and 20% below the mean respectively) while in the networks generated by the $M_{kg}$ and by $M_{er}$, they were considerably higher (19% above the mean). The networks generated by the other two models presented correlation values close to the mean. This fact is consistent with the degree variability of each model i.e., the larger is the range of the degree distribution generated by the model the lower is the correlation among centrality measures. The only metric that behaved differently was *eccentricity*. This happened due to a large number of tied centrality values in the $M_{er}$ networks. The lower the diameter of a network, the lower is the variety of distances among vertices and therefore the granularity of *eccentricity*. This fact penalizes its correlation values with all other centrality measures, granting it a 66% lower correlation than the average value over all networks.

The correlation among centrality measures varies, as expected, in different setups of parameters (including networks size) for a complex network model, but their variance were statically irrelevant. That is mainly because the chosen parameters preserved the fundamental characteristics of each complex model.

TABLE III. MEAN CORRELATION VALUES

| $C_c$ | | | | | | | |
|---|---|---|---|---|---|---|---|
| 0.75 | $C_b$ | | | | | | |
| 0.79 | 0.78 | $C_d$ | | | | | |
| **0.80** | 0.63 | 0.79 | $C_e$ | | | | |
| **0.83** | 0.77 | **0.91** | 0.79 | $C_i$ | | | |
| 0.76 | 0.61 | **0.80** | **0.94** | 0.77 | $C_s$ | | |
| 0.73 | **0.87** | **0.82** | 0.63 | 0.79 | 0.62 | $C_w$ | |
| 0.42 | 0.35 | 0.33 | 0.35 | 0.36 | 0.32 | 0.33 | $C_x$ |

### B. Centrality Measure Granularity

The second property of the measures that we analyzed was their granularity. We calculated the percentage of distinct centrality values for each metric in each network. For example, if we say that there are 50% distinct values in a network of 500 vertices, it means that there are 250 distinct (unique) centrality values for a given metric in that network. We considered six decimal places as the accuracy for the real value centralities as enough to distinguish the centrality values properly, considering the size of the networks used in our experiments.

The granularity property may be valuable in applications where centrality measures are used to differentiate/rank the vertices of the network [38] or as a heuristic for vertex selection and placement [25]. Tied values in these applications may suggest many unequal solutions for a given task in which only a best solution is required or desired. In addition, ties between centrality values of distinct vertices can be viewed as lack of information or incapability of the metric to differentiate the vertices properly, considering the fact that they are definitely unique in many domains. Moreover, correlation analysis between centrality measures and other domain-specific metrics are common. Such methods are impacted by equivalent values, reducing their accuracy and distorting the analysis.

Table IV presents the mean percentage of distinct values with 99% confidence intervals for each metric, grouped by the networks generated by the complex network models and by the non-isomorphic networks. Our results show that the granularity of all centrality measures presented nearly no difference in the expected values (considering the confidence intervals) in all networks generated by the complex network models. This is why we put them altogether in Table IV.

TABLE IV. MEAN PERCENTAGE OF DISTINCT VALUES

| Metric | Complex Models (%) | Non-Isomorphic (%) |
|---|---|---|
| $C_b$ | 98.72 ± 0.11 | 59.13 ± 1.72 |
| $C_c$ | 39.34 ± 0.60 | 55.07 ± 1.45 |
| $C_d$ | 19.11 ± 0.51 | 48.93 ± 1.12 |
| $C_x$ | 1.36 ± 0.04 | 26.86 ± 0.83 |
| $C_e$ | 99.57 ± 0.03 | 70.54 ± 1.92 |
| $C_i$ | 99.84 ± 0.02 | 69.36 ± 1.90 |
| $C_s$ | 94.15 ± 0.26 | 69.19 ± 2.00 |
| $C_w$ | 99.68 ± 0.04 | 69.56 ± 1.87 |

*Eccentricity* underperforms all other metrics with a high number of tied vertices due to its simple formulation (check Table I). It is followed by *degree* and then by *closeness* centralities in all synthetic networks. These three measures are by far worse than the others are in distinguishing vertices by their structural properties.

Nonetheless, *walk betweenness*, *information* and *eigenvector* are the metrics with the best granularity. They are followed narrowly by *betweenness* and then by *subgraph* centralities, forming a group with five high granularity metrics. This fact that is supported by their more complex formulations compared to the other three metrics.

Our results provide additional evidence that the *degree* measure is less fine-grained than *closeness* and both are inferior to *betweenness* in this aspect, as Freeman [17] originally thought but did not presented empirical results as we did. The intuition behind this fact is mainly because the simpler way that these three metrics are evaluated (please see Table I). Another interesting aspect of centralities analyzed in our experiments was the number of times each metric achieved the best-known granularity solution among the metrics tested. This information shows the number of times that one metric is better than all others are in distinguishing the vertices of a given network. These results are summarized in Table V.

TABLE V. PERCENTAGE OF TIMES WITH BEST GRANULARITY

| Metric | $N_{ni}$ | $M_{cs}$ | $M_{sf}$ | $M_{sw}$ | $M_{gr}$ | $M_{er}$ | $M_{kg}$ |
|---|---|---|---|---|---|---|---|
| $C_b$ | 38.8% | 97.6% | 62.8% | 78.3% | 70.2% | **100%** | 10.9% |
| $C_c$ | 33.7% | 0.0% | 0.0% | 0.0% | 0.0% | 0.0% | 0.0% |
| $C_d$ | 21.5% | 0.0% | 0.0% | 0.0% | 0.0% | 0.0% | 0.0% |
| $C_x$ | 4.9% | 0.0% | 0.0% | 0.0% | 0.0% | 0.0% | 0.0% |
| $C_e$ | **98.4%** | 87.7% | 51.7% | 47.9% | 62.0% | 55.0% | 82.5% |
| $C_i$ | 90.6% | 98.0% | 92.8% | 90.6% | **98.8%** | 67.3% | **94.8%** |
| $C_s$ | 93.4% | 32.4% | 31.3% | 34.2% | 34.3% | 32.0% | 21.3% |
| $C_w$ | 92.0% | **99.8%** | **100%** | **99.9%** | 76.5% | **100%** | 38.6% |

The cells highlighted in bold present the highest values. Notice that the columns total is higher than 100% because in many networks more than one metric achieves the best/top granularity performance. Table V reinforces even more the disparity in metrics granularity. The top overall metric *walk betweenness* is better in this aspect than all others are in most networks, excluding geographic and Kronecker networks, where *information* centrality performs well, and in non-isomorphic networks, where *eigenvector* is the best-qualified one. It also became evident the poor granularity of *eccentricity*, *degree* and *closeness* centralities compared to the others. The only networks where their granularity is at least equivalent to the other metrics were special cases of non-isomorphic networks, such as, the complete graph, when all centrality measures evaluate all vertices as being equally important.

V. CONCLUSIONS AND FURTHER WORK

The increasing availability of data on large technological, biological and social networks and the variability of applications have contributed to the development of many centrality measures. Nonetheless, little is known and there is a paucity of results about their properties and proper application. Most works in the area have focused on showing that each metric is distinct from the others and have not investigated their relationships. Thus, our work contributes towards filling this relevant gap. We do so by studying properties of eight centrality measures through a series of experiments with a large number of networks, generated with six complex network models and a set of non-isomorphic networks.

We revealed that five metrics (betweenness, eigenvector, information, subgraph and walk betweenness) outmatch the others (closeness, degree and eccentricity) in granularity. In addition, all metrics presented a high redundancy in their evaluation of vertex centrality. The pairs of metrics closer to each other in decreasing mean correlation value order (considering all networks) were: *eigenvector* and *subgraph* (0.94), *degree* and *information* (0.91), *betweenness* and *walk betweenness* (0.87), *closeness* and *information* (0.83). It is noticeable that *eccentricity* lags behind, with correlation values below 0.45 in the great majority of networks. This suggests that the application of centrality measures simultaneously can lead to limited or similar results.

Furthermore, we showed that the structural properties of the networks do not significantly affect the centralities granularity while they do have a considerable impact on their correlation. The analysis of the networks structural properties is relevant to a number of applications, including, e.g. multiagent scenarios [2][3]. Measures of complex network analysis, such as vertex centrality, have the potential to provide useful knowledge about patterns and behaviors in social networks. Another important contribution of this paper is to provide knowledge about centrality measures that helps in their selection and proper use in a given application domain. Further research includes investigation of network properties and their relationship with centrality measures' correlation, the application and comparison between metrics in directed and weighted networks, and the analysis of parametric measures properties and behavior.

ACKNOWLEDGMENT

This research is partly supported by the Brazilian Research Council CNPq and by the CAPES Foundation.

REFERENCES

[1] Adah, S.; Lu, X.; Magdon-Ismail, M.. Deconstructing centrality: thinking locally and ranking globally in networks. "ASONAM'13", Canada, 2013.

[2] Aadithya, K.V.; Ravindran, B. Game Theoretic Network Centrality: Exact Formulas and Efficient Algorithms (Extended Abstract). "AAMAS'10". 1459-1460, IFAAMAS, 2010.

[3] Araujo, R.M.; Lamb, L.C. Towards understanding the role of learning models in the dynamics of the minority game. "ICTAI'4". 727-731, 2004.

[4] Baig, M. B.; Akoglu, L.. Correlation of node importance measures: an empirical study through graph robustness. "WWW'15". Companion Volume: 275-281, 2015.

[5] Barabási, A.; Albert, R. Emergence of scaling in random networks. "Science". 286:509-512, 1999.

[6] Bei, X.; Chen, W.; Teng, S.-H.; Zhang, J.; Zhu, J.. Bounded budget betweenness centrality game for strategic network formations. "Theor. Comp. Sci". 412: 7147-7168, 2011.

[7] Bonacich, P.. Some unique properties of eigenvector centrality. "Social Networks". 29:555-564, 2007.

[8] Borgatti, S. P.; Carley, K.; Krackhardt, D. On the robustness of centrality measures under conditions of imperfect data. "Social Networks". 28:124-136, 2006.

[9] Butts, C. T. Exact bounds for degree centralization. "Social Networks". 28:283-296, 2006.

[10] Chakrabarti, D.; Faloutsos, C.. Graph mining: laws, generators, and algorithms. "ACM Computing Surveys". 38, 2006.

[11] Cohen, E.; Delling, D.; Pajor, T.; Werneck, R. F. Computing classic closeness centrality, at scale. "COSN'14". 2014.

[12] Costa, L. F.; Rodrigues, F. A.; Travieso, G.; Villas Boas, P. R.. Characterization of complex networks: a survey of measurements. "Adv.s in Physics". 56:167-242, 2008.

[13] Danowski, J. A.; Cepela, N.. Automatic mapping of social networks of actors from text Corpora: time series analysis. "Data Mining for Social Network Data". 12:31-46, 2010.

[14] Easley, D.; Kleinberg, J. "Networks, crowds, and markets: reasoning about a highly connected world". Cambridge U. Press, 2010.

[15] Erdős, P.; Rényi, A.. On random graphs I. "Publicationes Mathematicae". 6:290-297, 1959.

[16] Estrada, E.; Rodríguez-Velázquez, J. A.. Subgraph centrality in complex networks. "Physical Review E". 71:056103, 2005.

[17] Freeman, L. C.. "Centrality in social networks: conceptual clarification". Social Networks, 1:215-239, 1978/79.

[18] Goh, K. –I.; Oh, E.; Kahng, B.; Kim, D.. Betweenness centrality correlation in social networks. "Physical Review E". 67:017101, 2003.

[19] Hage, P.; Harary, F.. Eccentricity and centrality in networks. "Social Networks". 17:57-63, 1995.

[20] Hua, G.; Sun, Y.; Haughton, D.. Network analysis of US Air transportation network. "Data Mining for Social Network Data". 12:75-89, 2010.

[21] Jacomy M.; Venturini T.; Heymann S.; Bastian M.. ForceAtlas2, a continuous graph layout algorithm for handy network visualization designed for the Gephi software. "PLoS ONE". 9(6), 2014.

[22] Kaza, S.; Chen, H.. Identifying high-status vertices in knowledge networks. "Data Mining for Social Network Data". 12:91-107, 2010.

[23] König, M. D.; Tessone, C. J.; Zenou, Y.. Nestedness in networks: a theoretical model and some applications. "Theoretical Economics". 9:695-752, 2014.

[24] Leskovec, J.; Chakrabarti, D.; Kleinberg, J.; Faloutsos, C.; Ghahramani, Z. Kronecker graphs: an approach to modeling networks. "Journal of Machine Learning Research". 11:985-1042, 2010.

[25] Marchant, J.; Griffiths, N.; Leeke, M. Manipulating conventions in a particle-based topology. "COIN Workshop". 2015.

[26] Newman, M. E. J.. A measure of betweenness centrality based on random walks. "Social Networks". 27:39-54, 2005.

[27] Newman, M. E. J.. Scientific collaboration networks. II. Shortest paths, weighted networks, and centrality. "Physical Review E". 64:016132, 2001.

[28] Newman, M. E. J.. The structure and function of complex networks. "SIAM Review". 45(2):167-256, 2003.

[29] Newman, M. E. J.; Girvan, M. Finding and evaluating community structure in networks. "Physical Review E". 69(2):026113, 2004.

[30] Newman, M. E. J.; Park, J. Why social networks are different from other types of networks. "Physical Review E". 68:036122, 2003.

[31] Noble, D.; Grando, F.; Araújo, R. M.; Lamb, L. C.. The impact of centrality on individual and collective performance in social problem-solving systems. "GECCO'15". 2015.

[32] Şimşek, Ö.; Barto, A. G.. Skill characterization based on betweenness. "NIPS'9". 22:1-8, 2009.

[33] Soundarajan, S.; Eliassi-Rad, T.; Gallagher, B.. A guide to selecting a network similarity method. In "SDM". 1037-1045, 2014.

[34] Stephenson, K.; Zelen, M.. Rethinking centrality: methods and examples. "Social Networks". 11:1-37, 1989.

[35] Valente, T. W.; Coronges, K.; Lakon, C.; Costenbader, E.. How correlated are network centrality measures? "Connect (Tor)". 28(1):16-26, 2008.

[36] Watts, D. J.; Strogatz, S. H.. Collective dynamics of 'small-world' networks. "Nature". 393(6684):440-442, 1998.

[37] Xu, Y.; Hu, X.; Li, Y.; Li, D.; Yang, M.. Using complex network effects for communication decisions in large multi-robot teams. "AAMAS'14", 685-692, 2014.

[38] Żak, B.; Zbieg, A.. Heuristic for network coverage optimization applied in finding organizational change agents. "ENIC'14". 118-122, 2014.